\documentclass[traditabstract]{aa}
\usepackage{txfonts}
\usepackage{natbib}
\usepackage{graphicx}
\usepackage{rotating}
\usepackage[]{morefloats}
\newcommand{\Msun}{M$_{\odot}$}

\newcommand{\xsh}{X-Shooter }
\newcommand{\Lsun}{L$_{\odot}$}

\newcommand{\Mstar}{M$_{\star}$\ }

\newcommand{\Rsun}{R$_{\odot}$}

\newcommand{\Lk}{LkCa 15 }

\begin{document}

\title{Spectro-astrometry of LkCa 15 with X-Shooter: Searching for emission from LkCa 15b\thanks{Based on Observations collected with X-shooter at the Very Large Telescope on Cerro Paranal (Chile), 
operated by the European Southern Observatory (ESO). Program ID: 288.C-5013(A).}}


\author{Whelan, E.T. \inst{1}
  \and 
  Hu{\'e}lamo, N. \inst{2}
  \and
 Alcal{\'a}, J.M.  \inst{3}
 \and
  Lillo-Box, J.  \inst{2}
  \and
 Bouy, H.  \inst{2}
 \and 
 Barrado, D.  \inst{2}
 \and 
 Bouvier, J. \inst{4, 5}
\and
 Mer{\'i}n, B.  \inst{6}}

\institute{Institut f{\"u}r Astronomie und Astrophysik, Kepler Center for Astro and Particle Physics, Eberhard Karls Universit{\"a}t,  72076 T{\"u}bingen, Germany 
 \and  
 Centro de Astrobiolog{\'i}a, INTA-CSIC, Depto Astrof{\'i}sica, European Space Astronomy Cente (ESAC) Campus, PO Box 78, 28691, Villanueva de la Ca\~{n}ada, Madrid, Spain 
 \and 
INAF-Osservatorio Astronomico di Capodimonte, via Moiariello, 16, I-80131, Napoli, Italy 
\and
Univ. Grenoble Alpes, IPAG, F-38000 Grenoble, France
\and
CNRS, IPAG, F-38000 Grenoble, France
\and 
European Space Astronomy Center (ESAC), P.O. box 78, 28691, Villanueva de la Cañada, Madrid, Spain}

\titlerunning{Spectro-astrometry of LkCa 15 with X-Shooter} 
\date{}


  



\abstract {Planet formation is one explanation for the partial clearing of dust observed in the disks of some T Tauri stars. Indeed studies using 
state-of-the-art high angular resolution techniques have very recently begun to observe planetary companions in these so-called transitional disks. The goal of this work is to use spectra of the transitional disk object LkCa 15 obtained with X-Shooter on the Very Large Telescope to investigate the possibility of using 
spectro-astrometry to detect planetary companions to T Tauri stars. It is argued that an accreting planet should contribute to the total emission of accretion tracers such as H$\alpha$ and therefore
planetary companions could be detected with spectro-astrometry in the same way as it has been used to detect stellar companions to young stars. A probable planetary-mass companion was 
recently detected in the disk of LkCa 15. Therefore, it is an ideal target for this pilot study.  
We studied several key accretion lines in the wavelength range 300 nm to 2.2 $\mu$m with spectro-astrometry. While no spectro-astrometric signal is measured for any emission lines the accuracy achieved in the technique is used to place an upper limit on the 
contribution of the planet to the flux of the H$\alpha$, Pa$\gamma$, and Pa$\beta$ lines. The derived upper limits on the flux allows an upper limit of the mass accretion rate, log($\dot{M}_{acc}$) = -8.9 to -9.3 for the mass of the companion between 6 M$_{Jup}$ and 15 M$_{Jup}$, respectively, to be estimated (with some assumptions).}



\maketitle


\begin{figure*}
\centering
   \includegraphics[width=17cm, trim= 0cm 11cm 0cm 11cm, clip=true]{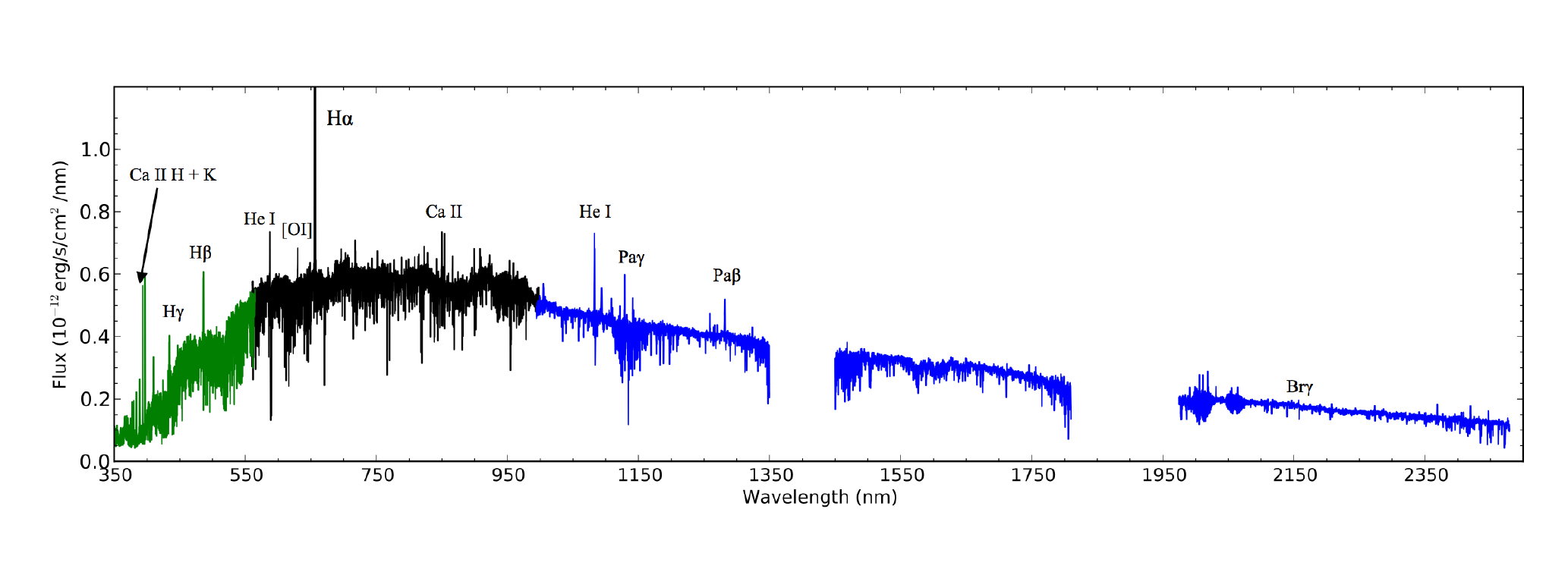}
     \caption{Full X-Shooter spectrum of  \Lk from the epoch 1 observations. The UVB arm is plotted in green, the VIS arm in black and the NIR arm in blue. 
     Major features are marked. The spectra as shown here have not been extinction corrected.}
  \label{spec_full}     
\end{figure*}

\section{Introduction}
Transitional disks (TDs) around young stellar objects (YSOs) can be defined as accretion disks with an inner region that is significantly lacking in dust. Grain growth and clearing by a planet are two possible explanations for the removal of dusty material from an accretion disk \citep{Alexander14}. Therefore, YSOs with TDs offer an interesting opportunity to study planet formation at the earliest stages \citep{Huelamo13}. 
TDs are characterised by a lack of mid-infrared (IR)
 emission and a rise into the far-IR \citep{Merin10}. The clearing of material can be discussed in terms of a hole or a gap. A TD is described as having a gap when both an inner and an outer disk is present and a hole when there is no inner disk.  Possible planetary companions have very 
recently been identified in the TDs of two T Tauri stars (TTSs), T Cha and LkCa 15 \citep{Huelamo11, Kraus12}. Both T Cha and LkCa 15 were selected for these
 studies as their spectral energy distributions (SEDs) showed no evidence of dust in their inner disks. The dust could have been cleared by a planetary mass object \citep{Brown07, Esp07} and these detections were done using aperture masking techniques \citep{Tuthill10}.

Much can be learned about TDs and planet formation through the study of the accretion properties of YSOs with TDs. For example, measurements of the mass accretion rate ($\dot{M}_{acc}$) of a sample of TDs by \cite{Manara14} have shown $\dot{M}_{acc}$ to be comparable to $\dot{M}_{acc}$ in Classical 
TTSs (CTTSs). CTTS are Class II low-mass YSO with full accretion disks and strong accretion rates ($\dot{M}_{acc}$ $>$ 10$^{-9}$ \Msun / yr; Muzerolle et al. 2003). The TD phase of a TTS is postulated to be intermediate between the CTT phase and the Class III phase where the disk is mostly dissipated and accretion has stopped.  The results of \cite{Manara14} reveal information about the gas content of TDs in the regions close to the star and imply that these regions are gas-rich. 
As planet formation models also predict accretion onto recently formed planets 
investigations of $\dot{M}_{acc}$ in accreting objects at the planetary/brown dwarf boundary are also important for the theory of planet formation \citep{Zhou14}. For example, \cite{Lovelace11} explain that magnetospheric
processes may govern accretion onto young gas giants in the isolation phase of their development when they have
already cleared gaps in their surrounding disks. In this evolutionary stage they are expected to accrete slowly through their magnetic field lines like low-mass stars and it should be possible to detect strong emission in accretion indicators like the H$\alpha$. \cite{Owen14} postulate that the high mass accretion rates measured for YSOs with TDs offer a clue to the origin of the dust poor inner regions of transitions disks. It is argued that the observed accretion rates in TDs imply a high accretion luminosity originating from the forming planet of $\geq$ 10$^{-3}$ \Lsun. The models of \cite{Owen14} require that the planets accrete at least half of the material flowing into the gap in-order to trap sufficient sub-micron dust particles.

Here we report on X-Shooter observations of LkCa 15. 
Three epochs of X-Shooter data were obtained between December 2011 and March 2012 (Table 1). 
The primary objective of this project was to investigate the possibility of detecting the planetary companion discovered by \cite{Kraus12} using spectro-astrometry 
(SA) and thus to test if SA can be used to detect recently formed planets (see Section 2.2).  
The motivation to use this technique is that planets are formed very close to the central star and it is 
difficult to detect and spatially resolve them at optical wavelengths. 
This paper is divided as follows. Firstly, the target, the technique of SA, and the \xsh observations are described in Section 2.
In Section 3 the results of the spectro-astrometric analysis are outlined. 
In Section 4 the accretion properties of the star and companion are investigated and the results of the SA are used to estimate
upper limits to the flux of important accretion tracers, e.g. H$\alpha$, and thus upper limits to $\dot{M}_{acc}$.  The conclusions are given in Section 5 and further information on the SA and 
the artefacts which we found to be affecting the \xsh data are given in the Appendix.




\section{Target, spectro-astrometry, and observations}

\subsection{Target} 
\Lk (04$^{h}$39$^{m}$17$^{s}$.8, +22$^{\circ}$21$^{\arcmin}$03.$^{\arcsec}$2) located in the Taurus-Auriga star forming region (d = 145 $\pm$ 15~pc, Loinard et al. 2005) 
is a K2 TTS \citep{Kenyon95} with an estimated age of $\sim$ 2 Myr \citep{Kraus12}. Studies based on the analysis of the IR spectral energy distribution (SED) of \Lk and sub-millimeter observations,
showed it to have a TD with a mass of $\sim$ 55~M$_{Jup}$ and a gap of $\sim$ 50~AU \citep{Andrews05, Esp07, Andrews11}. \Lk also displays  
a  significant  near-infrared  excess \citep{Esp07}
and millimeter emission from small orbital radii \citep{Andrews11} 
which points towards the presence of an inner  AU-sized disk.
The mass of \Lk has been estimated from the rotation curve of the disk at 0.97 $\pm$ 
0.03~\Msun\ \citep{Simon00}. \cite{Manara14} estimate \Mstar at 1.24 $\pm$ 0.33~\Msun\ using the evolutionary models of \cite{Baraffe98}. \cite{Kraus12} reported 
the detection of a possible planetary mass companion (LkCa 15b) at a distance of $\sim$ 16~AU to 21~AU from LkCa 15, placing it at the centre of the previously detected gap. 
They estimate the mass of LkCa 15b at $\sim$ 6~M$_{Jup}$ to 15~M$_{Jup}$. Specifically \cite{Kraus12} observe a 
blue (K$^{´}$ or 2.1 $\mu$m) point source surrounded by co-orbital red (L$^{´}$ or 3.7 $\mu$m) emission that is resolved into two sources. They suggest that the most likely 
geometry is of a newly formed gas giant planet that is surrounded by dusty material, and which has been caught at its epoch of formation. They refer to the blue point source as the 
central source (CEN) and the red sources as the north-east (NE) and south-west (SW) sources. \cite{Ireland14} report a clockwise orbital motion of 
6.0 $\pm$ 1.5 degrees per year for the CEN source. Finally $\dot{M}_{acc}$ has been estimated for \Lk at 3.6 $\times$ 10$^{-9}$~
\Msun / yr$\sim$ \citep{Ingleby13} and 4 $\times$ 10$^{-9}$~\Msun / yr  \citep{Manara14}.



\begin{figure*}[!ht]
\centering
   \includegraphics[width=16cm, trim= 0cm 0cm 0cm 0cm, clip=true]{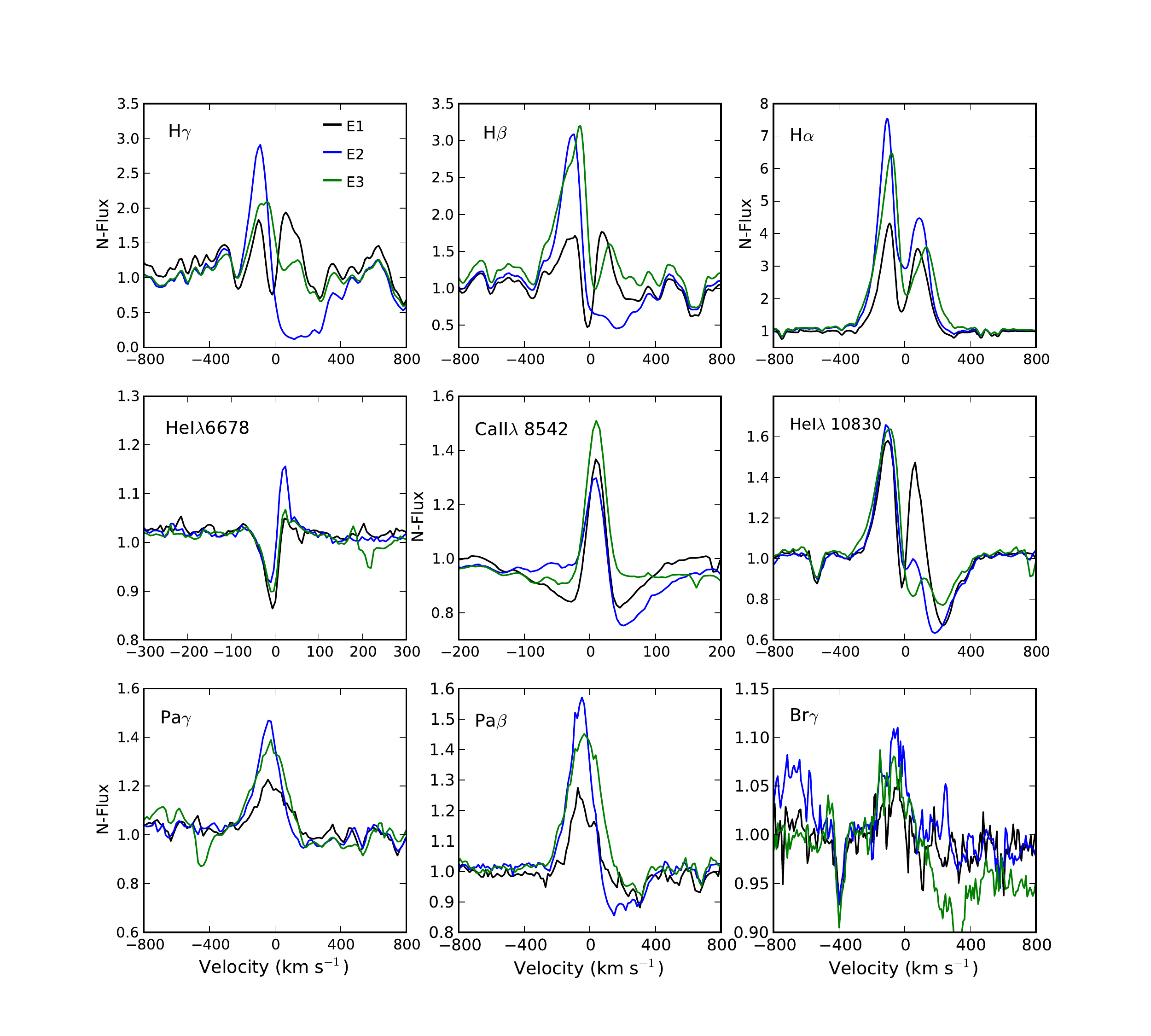}
     \caption{Key accretion tracers detected in the LkCa 15 X-Shooter spectrum.}
  \label{lines}     
\end{figure*}

\subsection{Spectro-astrometry}
SA is a technique by which spatial information with a high precision can be recovered from a simple seeing limited spectrum \citep{Whelan08, Bailey98}. 
Thus is it a very powerful method by which the limits placed on the spatial resolution of a spectroscopic observation by the seeing can be overcome. 
The technique is based on measuring the centroid of the spatial profile of an emission line region as a function of wavelength with respect to the continuum centroid. 
Typically this has been done using Gaussian fitting to produce what is referred to as a position spectrum.  
For Gaussian fitting the precision to which the centroid can be measured is given by 
\begin{equation}
\sigma = \frac{FWHM}{2.3548(\sqrt{N_{p})}}
\end{equation}
It is the fact that the precision of this technique is mainly reliant on the number of detected photons (N$_{p}$) or the signal to noise (S/N) ratio of the observation that makes it so useful. This statement assumes that observations of the source under investigation, where a large number of source photons are detected and where the detector and background noise is not significant, are easily achieved. Other considerations when doing a spectro-astrometric analysis are any curvature in the spectrum due to instrumental effects and the contribution of 
the continuum emission to the results. It is straight-forward to fit the position spectrum to remove the instrumental curvature before considering the final results. 
The continuum emission will contaminate the position spectrum in that any displacement in the emission line region will be reduced by a factor equal 
to (I$_{line}$ + I$_{cont}$) / I$_{line}$. Studies have dealt with the issue of continuum contamination by subtracting the continuum emission \citep{Whelan04} 
or by multiplying the position spectrum of the line (after the removal of instrumental curvature) by this factor \citep{Takami01}.

Some of the first uses of SA were for the detection of binaries that were not well separated and thus were difficult to directly detect \citep{Bailey98, Takami03} and for the detection of 
small scale jets in emission lines that were not traditionally believed to have a strong jet component \citep{Whelan04}. It has since been used in further interesting 
ways including to study Keplearian rotation in circumstellar disks \citep{Ponto08}, black holes \citep{Gnerucci13}, planetary nebulae \citep{Blanco14} and jets from 
brown dwarfs (BDs) \citep{Whelan05, Whelan09b}.  While SA only recovers spatial information along the position angle (PA) used for the slit, it is possible to do 2D 
spectro-astrometry and thus assign a proper on-sky direction to measured offsets. Firstly, this can be done by taking observations using orthogonal slit PAs 
and combining the offsets measured to estimate a direction and PA for the spatial offsets. This method has been adopted to assign directions and PAs to BD 
outflows \citep{Whelan14b, Whelan12}. Secondly, 2D SA can be acheived in a more direct way by analysing integral field spectra (IFS). The advantage of 
integral field spectra is that it is often procured using Adaptive Optics (AO) techniques and so due to the much reduced FWHM of the spatial profile an even better spectro-astrometric 
precision can be achieved. For example, \cite{Davies10} combined SA with IFS to detect the outflow from the massive YSO W33A with a precision of 0.1~mas.  

Finally, the topic of spectro-astrometric artefacts should also be discussed. It has been shown that false spectro-astrometric offsets can be introduced into an 
observation due to a distorted point spread function (PSF) caused by poor tracking or uneven slit illumination \citep{Brannigan06}. These artefacts can 
mimic the signature from a disk or jet. To check for spectro-astrometric artefacts one can obtain parallel spectra for example, at PAs of 0$^{\circ}$ and 
180$^{\circ}$. Any real signatures will be present at both PAs and their direction will be inverted between spectra \citep{Brannigan06, Podio08}. A 
further check is to analyse lines such as telluric absorption lines where one would not expect to see any signature \citep{Davies10, Whelan05}. It is possible to model and thus remove 
 these artefacts  \citep{Adams15, Podio08}.
  

\subsection{X-Shooter observations}

The primary datasets analysed here are X-Shooter observations. X-Shooter, a second generation instrument on the European Southern Observatory's 
Very Large Telescope (ESO VLT), covers a large wavelength range from the ultraviolet to the near-IR \citep{Vernet11} and is proven to be extrememly useful for studies 
of accretion in YSOs \citep{Alcala14, Whelan14a, Whelan14b}. Using X-Shooter in long-slit mode three epochs of data of LkCa 15 were obtained. \Lk was observed in nodding mode with a single node exposure time 
of 220~s for the ultraviolet (UVB) and visible (VIS) arms and 200~s for the near-IR (NIR) arm. This yielded a nominal exposure time of almost 30~mins after 
one ABBA cycle. The slit widths of the UVB, VIS and NIR arms were 0\farcs8, 0\farcs7 and 0\farcs9, respectively. This choice of slit widths resulted in spectral resolutions
 of 6200, 11000 and 5300 for each arm, respectively. The pixel scale is 0\farcs16 for the UVB and VIS arms and 0\farcs21 for the NIR arm. In each epoch two spectra were obtained. In the first epoch (E1) spectra were obtained parallel and anti-parallel to the PA of the \cite{Kraus12} CEN component which is assumed 
to be the PA of the companion. We take this PA to be 333$^{\circ}$. This value comes from the average of the PAs reported for the K band emission in Table 2 of \cite{Kraus12}.
The orbital motion reported in \cite{Ireland14} translates to a $\sim$ 8$^{\circ}$ change in the PA of the CEN source between the \cite{Kraus12} observations and the data presented here.  
Therefore Lk Ca15 b would be at $\sim$ 325$^{\circ}$ at the time of the \xsh observations and therefore would still lie within the slit in our observations. For E1 the purpose of the anti-parallel spectra was to check for spectro-astrometric artefacts. In epochs 2 and 3 (E2, E3) spectra were obtained parallel and perpendicular to the PA of the companion. The purpose here was to recover the PA of any spectro-astrometric signature. Taking one 
ABBA cycle as one spectrum a total of six spectra of \Lk were obtained. 
All this information is summarised in Table 1. 

The \xsh data reduction was performed independently for each arm using 
the X-Shooter pipeline version 1.3.7 \citep{Mod10}.\footnote{An
 improved pipeline running in a new environment \citep{Freudling13} has been released after the conclusion of our data reduction.
 However, we have conducted tests with the X-Shooter pipeline team at ESO and found that these improvements do not affect the
 results presented in this paper.} The pipeline provides 2-dimensional, bias-subtracted, flat-field corrected, order-merged, background-subtracted 
and wavelength-calibrated spectra.  Flux calibration was achieved within the pipeline using a response function derived from the spectra of flux-calibrated standard stars observed on the 
same night and of \Lk. Following the independent flux calibration of the \xsh arms,  the internal
consistency of the calibration was checked by plotting together the three spectra extracted
from the source position and visually examining the superposition of overlapping
spectral regions at the edge of each arm. Any misalignment was corrected for by scaling the UVB and NIR spectra to the VIS continuum level. To correct for slit losses the spectra were normalised to the average photometry available in the literature. An average V-band magnitude of 12.15 mag was adopted \citep{Grankin07}. The normalization was done by applying a correction factor equal to 1.8. The correction factor was calculated by comparing the spectroscopic flux, F$_{\rm SPEC}$(5483\,\AA)=2.8e-14\,erg/s/cm$^2$/\AA, with the photometric flux F$_{\rm phot}$(5483\,\AA)$=$F$_0$(5483\AA)$\cdot$10$^{-0.4\cdot12.15}=$5.1e-14\,erg/s/cm$^2$/\AA,where the zero magnitude flux F$_0$(5483\,\AA)$=$3.67e-9\,erg/s/cm$^2$/\AA \citep{Johnson64, Johnson65}. We have confirmed that the correction factor is the same for all the wavelengths where photometry is available (B \& R from \cite{Grankin07}, and J, H, \& K from the Two Micron All-Sky Survey (2MASS). Finally, the correction for the contribution of telluric bands was done using the telluric standards observed with the same instrumental set-up, and close in airmass to \Lk. More details on data reduction, as well as on flux calibration and correction for telluric bands can be found in \cite{Alcala14}. 

\begin{table}[h]
\begin{tabular}{cccccc}       
 \hline\hline 
Epoch      &Slit PA ($^\circ$)     &Date   &Time (UT)            &Seeing (\arcsec)   
 \\ 
\hline
1              &333                                             & 2011-12-02                    &03:52         &0.85           
\\
1              &{\bf153}                                             & 2011-12-02                    &04:22         &0.84     
\\ 
2              &333                                               &2012-02-27                     &00:24         &1.3    
 \\
2              &243                                             &2012-02-27                     &00:55       &1.5        
\\ 
3              &333                                               &2012-03-07                     &00:11        &1.15      
\\ 
3              &243                                             &2012-03-07                     &00:43         &1.35    
\\
 \hline  
\end{tabular}
\caption{X-Shooter Observations of \Lk. Each epoch contains two observations. In E1 data was taken parallel and anti-parallel to the companion PA. In E2 and E3 the slit was parallel 
and perpendicular to the companion PA. The seeing is for the VIS arm and it should be noted that the seeing can vary between the three arms. }
\end{table}

\section{Results}

\subsection{Spectro-astrometric analysis}
In Section 2 the various steps involved in the reduction of the \xsh data by the pipeline are outlined. The final product is a 2D spectrum where the sky has 
been subtracted. The spectro-astrometric technique is described in Section 2.2. The goal here was to investigate 
if it is feasible to detect LkCa15 b through the spectro-astrometric analysis of emission lines which could also be tracing this companion. The detection of the planet 
is theoretically possible for the same reason as it is possible to separate two close stars using SA \citep{Takami03}. If the planet emits in an accretion tracer, due to accretion onto it, 
one will see an offset in the position spectrum of the accretion line at any point in the line where the emission from the planet is not dominated by the emission from the star. 
The size of the offset ({\it S}) will depend on the relative strength of the line flux from the star ({$\it F_{*}$}) and planet ({$\it F_{pl}$}) and the separation between the star 
and its companion  ({\it D}). This assumes that the continuum emission has been removed (see Section 2.2).


  \begin{figure*}[!ht]
\centering
    \includegraphics[width=5.2cm]{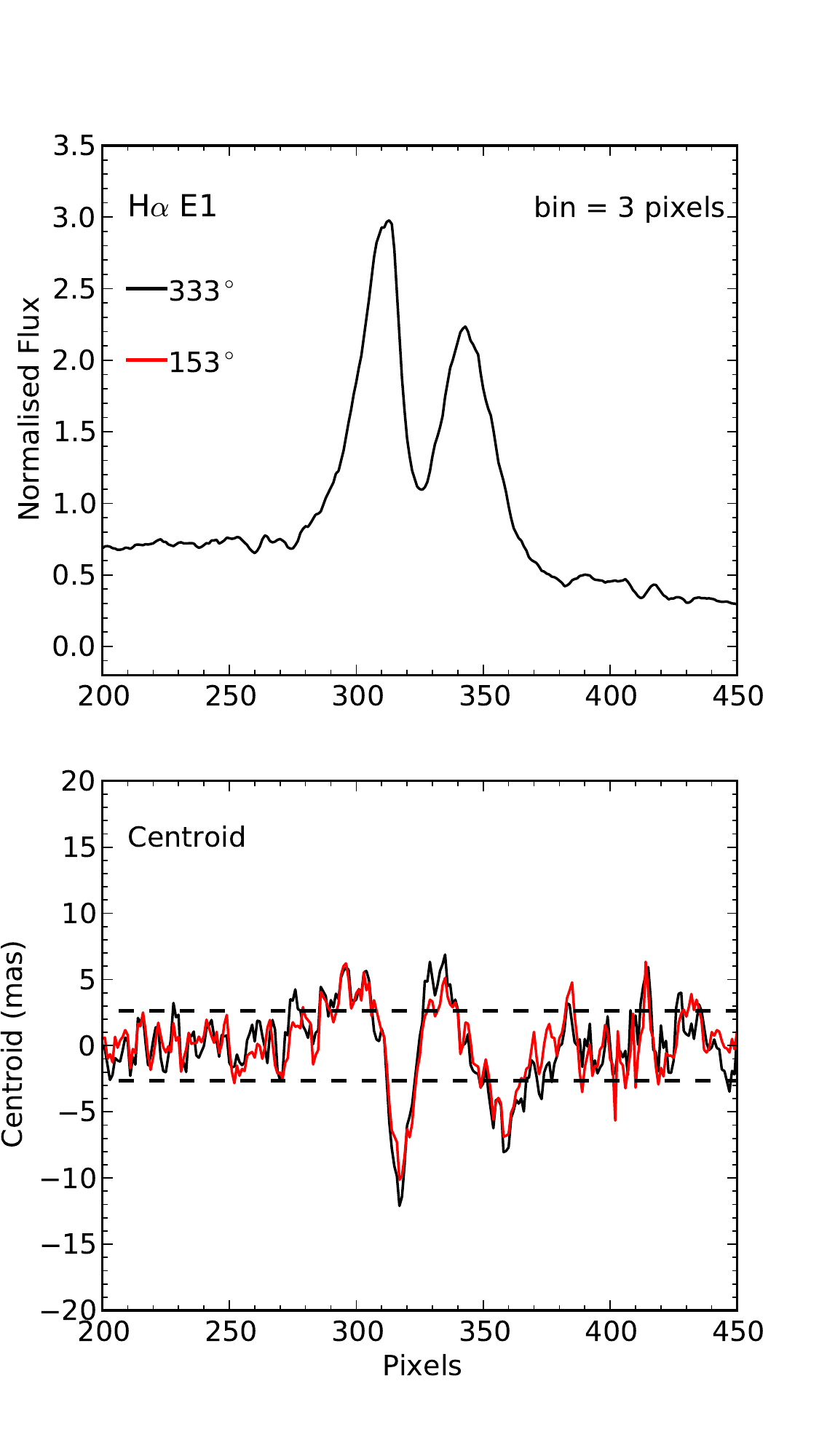}
    \includegraphics[width=5.2cm]{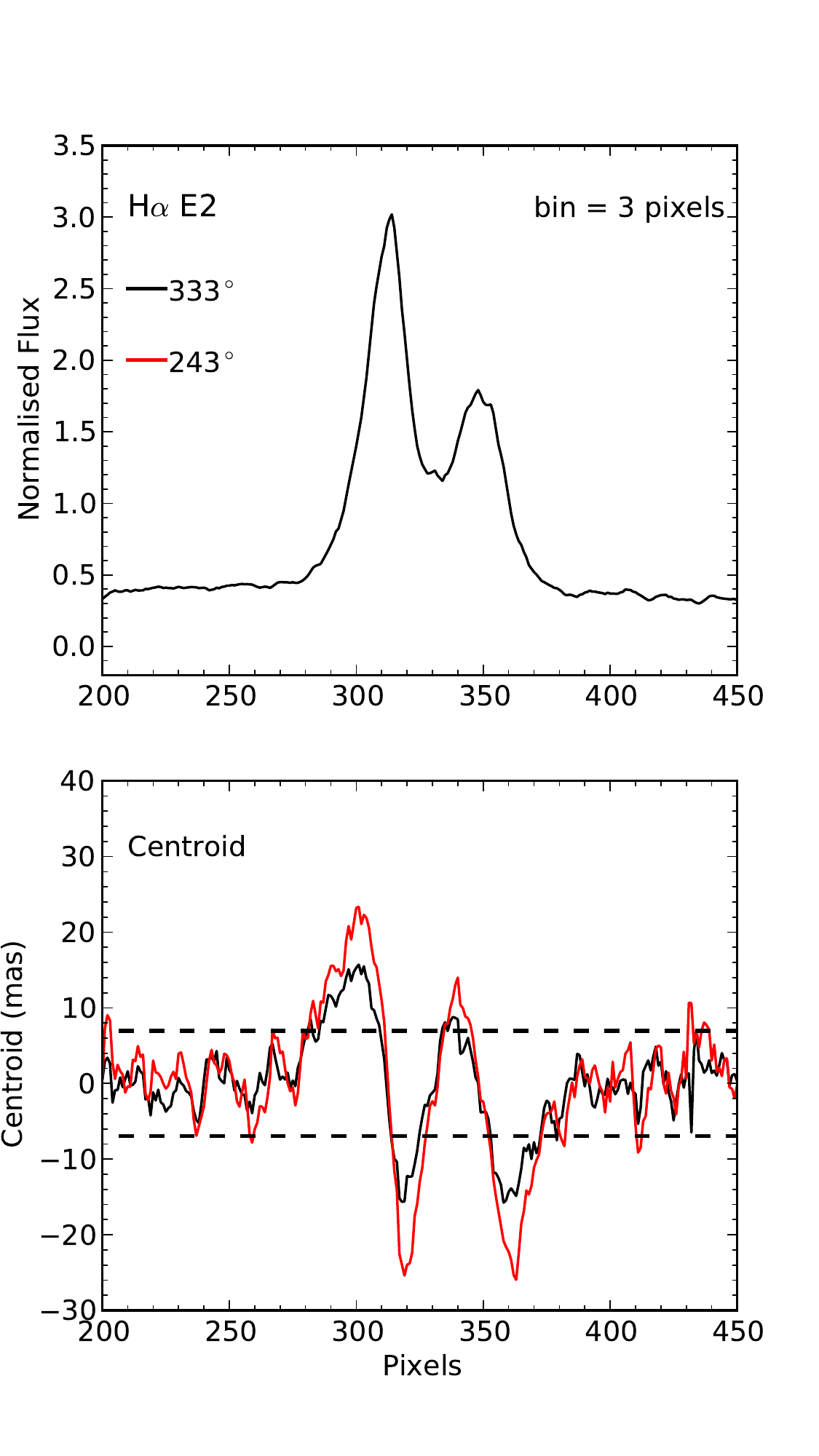}
     \includegraphics[width=5.2cm]{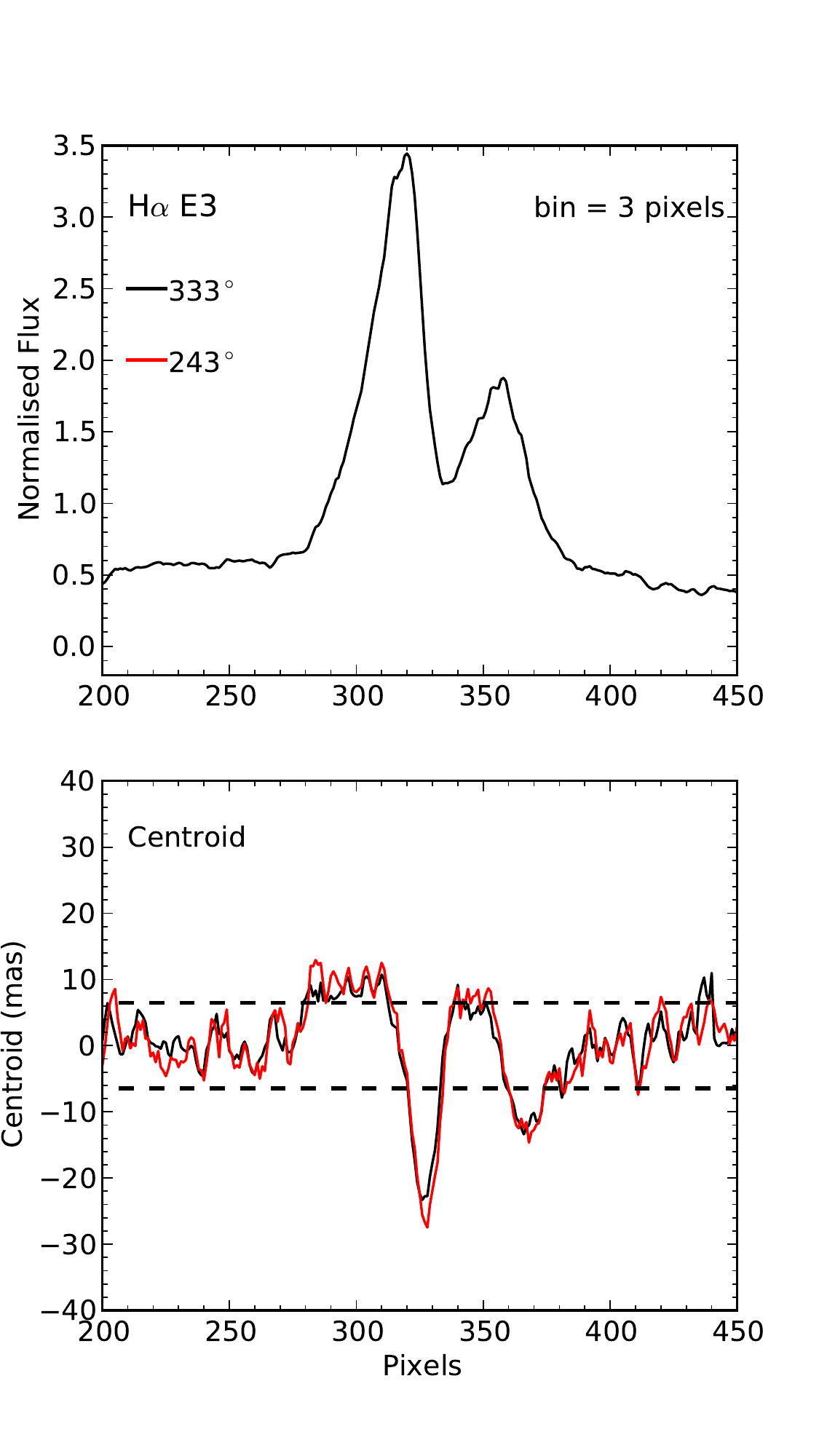}
     \caption{Spectro-astrometric analysis of the H$\alpha$ line in each epoch. What is shown here are the normalised intensity profiles and the position spectra. The size of the bin used in the analysis is given. The black line is the data along the estimated companion PA and the red line the data parallel and perpendicular to this PA. 
     The dashed horizontal lines are the 1-$\sigma$ uncertainty. }
  \label{SA_Ha}     
\end{figure*}

It was expected that the goal of detecting LkCa15~b would be 
challenging mainly due to the fact that any contribution to the accretion tracers from the planet would be small. Furthemore, emission from transitional disk objects in accretion tracers like H$\alpha$
is known to be variable and therefore the chances of detecting the companion would be greatly improved if the star is observed in a state where ({$\it F_{*}$}) is at a minimum. Having three epochs of data increased the chances of observing ({$\it F_{*}$}) at a minimum. It was found that a
Voigt fit provided a better fit to the wings of the spatial profile than the Gaussian fit. Hence, for the analysis of the \xsh data we use a Voigt fit at all times. The precision in the Gaussian fit 
calculated using Equation 1 is equivalent to the 1-$\sigma$ scatter in the position spectrum of the continuum emission. By comparing the 1-$\sigma$ scatter in the continuum  calculated using the Gaussian 
and Voigt fits it was shown that the precision in the Voigt fit can be well approximated by Equation 1. Finally, it should also be noted that the spectra were binned when doing SA to increase the accuracy. The bin size was three pixels.

The first step in our analysis was to apply SA to the UVB, VIS, and NIR spectra produced by the X-Shooter pipeline, in each epoch and at each PA. However, it was found that the UVB and VIS data was affected by a spatial aliasing problem which we found arose due to the rebinning of the spectra from the physical pixel
space (x,y) to the virtual pixels (wavelength, slit-scale). Thus the processed UVB and VIS data was not suitable for spectro-astromertic anaysis. This problem is important for any future spectro-astrometric studies with X-Shooter and therefore it is further disucssed in the Appendix. We proceeded with the our analysis by fitting the UVB and VIS spectra produced by the pipeline before the wavelength calibration step. Thus these spectra were bias subtracted, flat-fielded and any field cosmetics were removed. 
For the NIR arm the wavelength and flux calibrated spectra were used.

We investigated the most prominent accretion tracers found across the three arms i.e H$\gamma$, H$\beta$, H$\alpha$, HeI$\lambda$6678, CaII triplet, HeI 1.083$\mu$m, Pa$\gamma$ and Pa$\beta$. 
Please refer to \cite{Alcala14} for a further description of these accretion tracers. In Figure \ref{lines} these accretion tracers are shown. The emission lines in each epoch and PA are presented and it is clear that the variability between the two PAs is not significant but is significant between the different epochs. A study of the variability of LkCa 15 is beyond the scope of this paper and a second paper is in preparation on this subject. Position spectra were prepared for all the lines shown in Figure \ref{lines} and no spectro-astrometric offset was detected for any of these lines.
The results of the analysis of the H$\alpha$ and NIR Pa$\gamma$, Pa$\beta$, Br$\gamma$ lines are now discussed in more depth. We chose to focus on the H$\alpha$, Pa$\gamma$ and Pa$\beta$ lines as they are the best choice for estimating M$_{acc}$ in the star and planet. This is due to the fact that they are the lines which are least affected by absorption. Furthermore, the H$\alpha$ line is the line in which we are most likely to detect the planet \citep{Close14}.  The Br$\gamma$ line is included as it is relevant to the results of \cite{Kraus12}.

\begin{figure}
\centering
\includegraphics[width=9cm, trim= 1cm 9cm 1cm 9cm, clip=true]{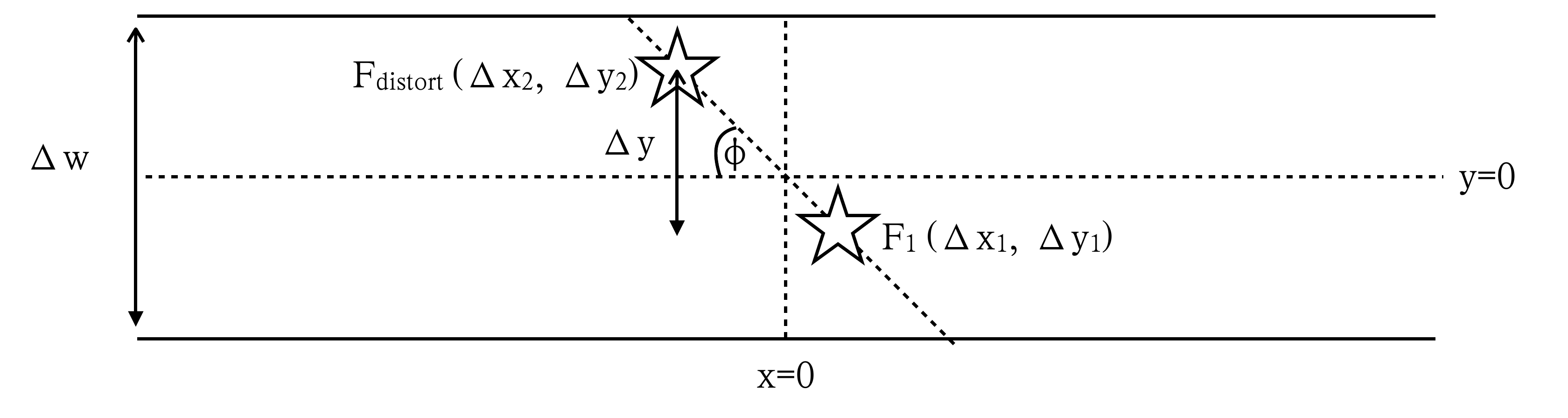}
     \caption{Geometry used for the simulations of the artefacts. The ratio between F$_{distort}$ and F$_{1}$ is kept constant at 1 for all the simulations while $\Delta$y and $\phi$ is varied. }
  \label{slit}     
\end{figure}

\subsection{Analysis of the H$\alpha$ line}

\begin{figure*}[ht!]
\centering
  \includegraphics[width=5cm]{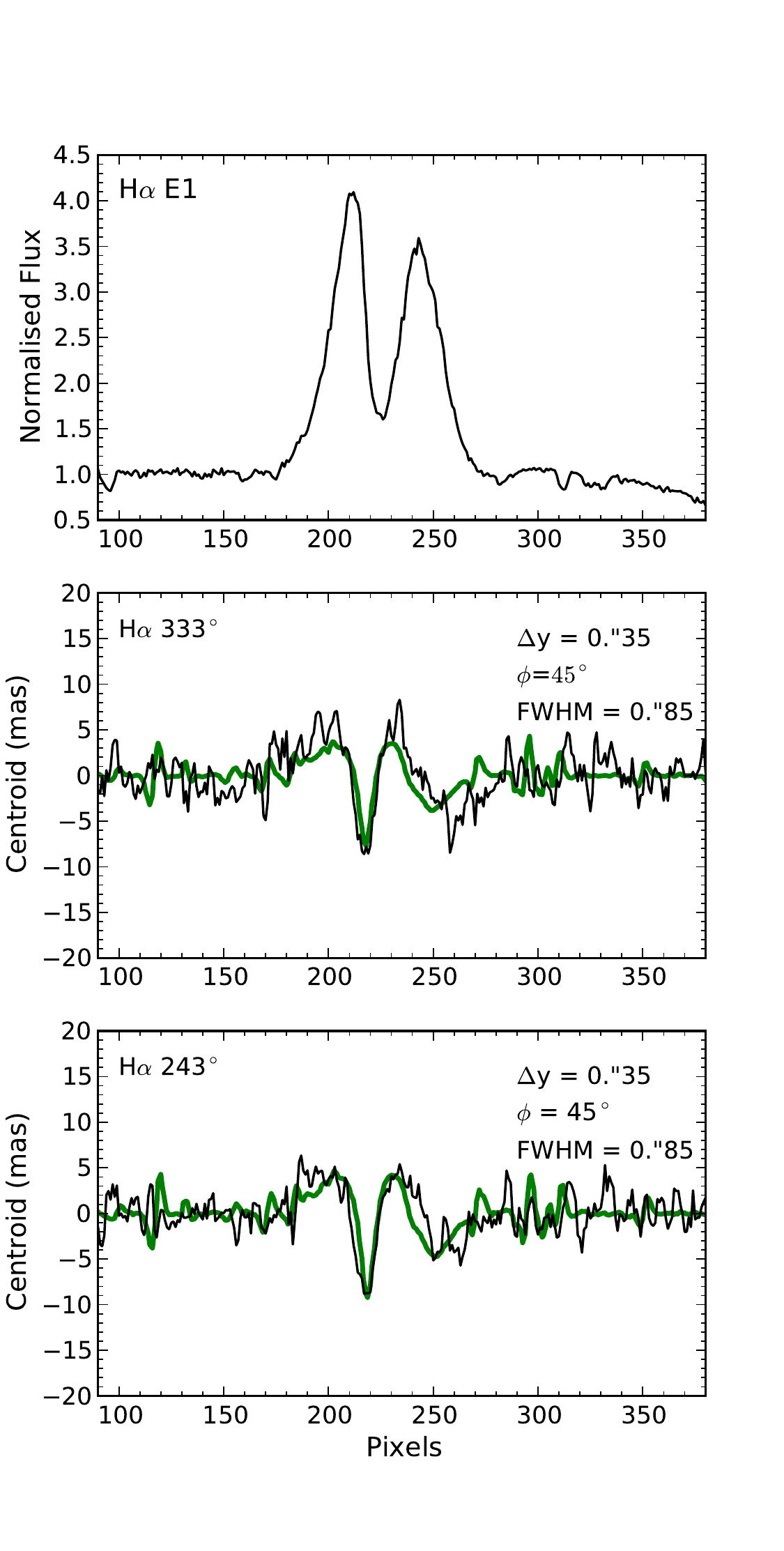}
  \includegraphics[width=5cm]{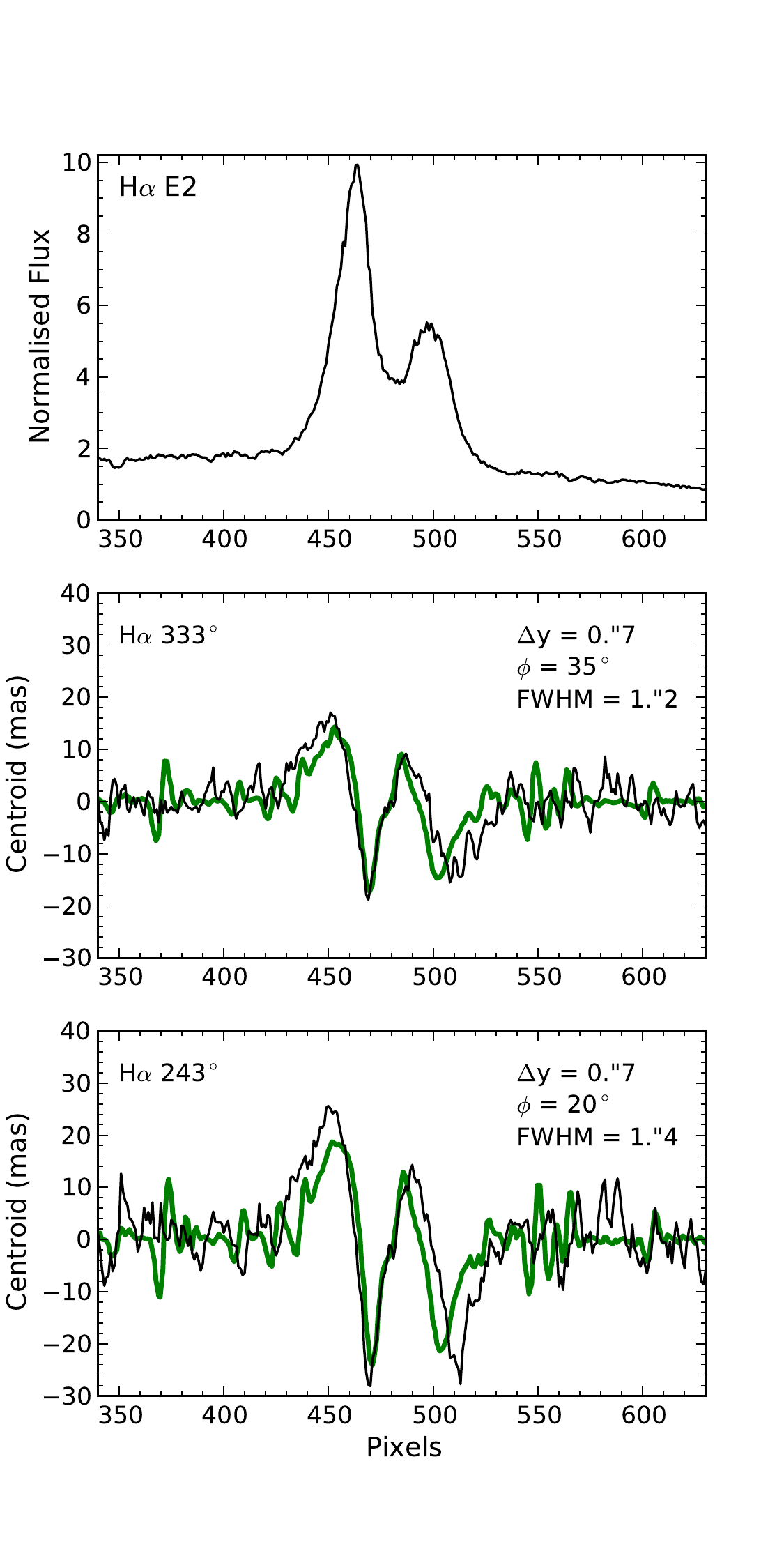}
  \includegraphics[width=5cm]{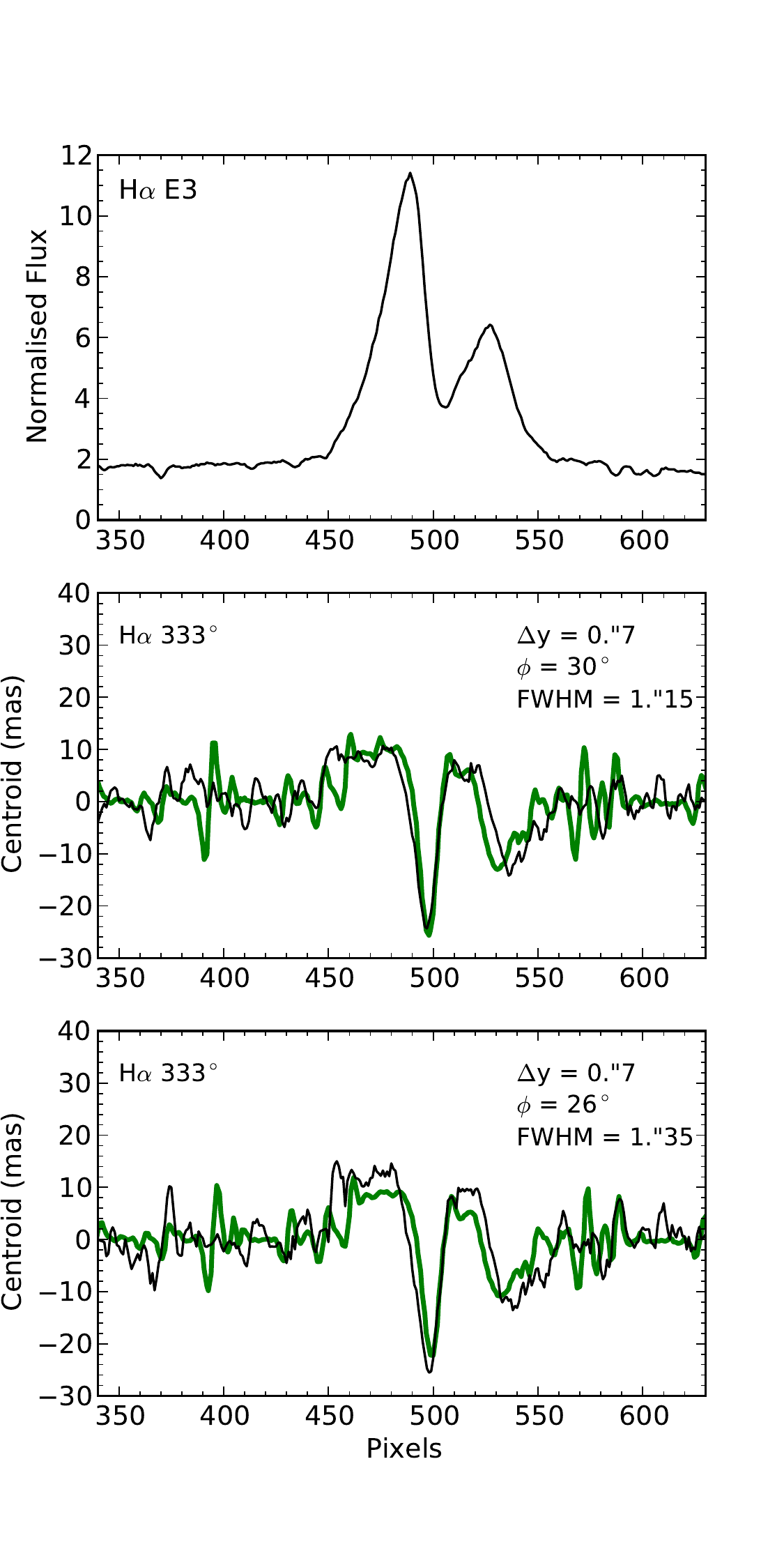}
      \caption{The spectro-astrometric artefacts detected in the raw data are simulated using equations 2 to 5. The green line is the simulated position spectrum.}
  \label{Ha_model}     
\end{figure*}

Due to the artefacts introduced by the wavelength calibration of the \xsh data the spectra before wavelength calibration was used for the analysis of the H$\alpha$ line. Strong signatures were detected in all epochs and PAs and  the results are shown in Figure \ref{SA_Ha}. In E1 
where slits parallel and anti-parallel to the companion PA were used the offsets do not change with PA. If this signature were real it would be expected that the direction of the offset would change between the two slit PAs \citep{Brannigan06}. Therefore, what is being seen here is an artefact. Using the approach of \cite{Brannigan06} the artefacts seen in the H$\alpha$ line are modelled as being due to a distorted PSF caused by uneven illumination of the slit. Given a single point source with flux F$_{1}$ located within the slit, a distortion in the PSF can be simulated by placing a percentage of the flux of the point source (F$_{distort}$) at a new position off centre of the slit. This geometry is illustrated in Fig. \ref{slit}. The centroidal position as a funcion of wavelength for this PSF is then given by 
\begin{equation}
X_{cent} = \frac{\int \int xF(x,y) f(\lambda + \Delta \lambda(y)) dy dy}{\int \int F(x,y) f(\lambda + \Delta \lambda(y)) dy dy}
\end{equation}
where
\begin{equation}
F(x,y) = G_{1}(x,y) + G_{1}(x,y)
\end{equation}
and
\begin{equation}
G_{1}(x,y) = \frac{2.773}{(FWHM)^{2}\pi}F_{1} e^{-2.773\frac{(x-\Delta x_{1})^{2}+(y-\Delta y_{1})^{2}}{(FWHM)^{2}}}
\end{equation}
\begin{equation}
G_{2}(x,y) = \frac{2.773}{(FWHM)^{2}\pi}F_{distort} e^{-2.773\frac{(x-\Delta x_{2})^{2}+(y-\Delta y_{2})^{2}}{(FWHM)^{2}}}
\end{equation}

\noindent In the above equations $\Delta$x$_{1}$, $\Delta$y$_{1}$ and $\Delta$x$_{2}$, $\Delta$y$_{2}$ are the positions of F$_{1}$ and F$_{distort}$ with respect to the centre of the slit. The shape of the signatures produced by the distortion in the PSF depend on the spectrum of the star while the magnitude of the false offsets depends on the the ratio between F$_{1}$ and F$_{distort}$, the seeing (FWHM) and the positions $\Delta$x$_{1}$, $\Delta$y$_{1}$ and $\Delta$x$_{2}$, $\Delta$y$_{2}$. For their simulations \cite{Brannigan06} take a value for the PA of the distortion ($\phi$) of 45$^{\circ}$ (See Figure 4). Their results showed that for constant values of F$_{1}$:F$_{distort}$ and FWHM, the magnitude of the false offsets increased with increasing separation (shown as $\Delta$y in Fig. \ref{slit}) between F$_{1}$ and F$_{distort}$.

In Fig. \ref{Ha_model} the results of the simulations of the \xsh H$\alpha$ position spectra are shown. For all epochs we set F$_{1}$  equal to F$_{distort}$ and it was found that in order to reproduce the measured artefacts it was necessary to not only vary $\Delta$y between epochs but also $\phi$. Thus the PSF distortions present in our data are particularly strong and vary significantly between the three epochs. Interestingly our results show that the size of the false offsets increased even as the seeing increases (see Table 1). The slit width remained constant at 0\farcs7. The full extent of the offsets increases from $\sim$ 15~mas in E1 to $\sim$ 55~mas in E2 and $\sim$ 35~mas in E3. \cite{Brannigan06} demonstrated that the size of the artefacts could be minimsed by increasing the FWHM with respect to the slit width (while keeping the parameters describing the PSF constant). They suggest that using a slitwidth which is narrow in comparison to the seeing could offer some protection against these artefacts. However, our results for E2 and E3 show that this will not solve the problem with artefacts in cases of a strong distortion of the PSF.



To check for a residual spectro-astrometric signature we subtracted the artefacts firstly by subtracting the E1 anti-parallel spectra and secondly by using the model results. No real spectro-astrometric signature was seen. Although no spectro-astrometric signature from the planet is detected the uncertainty on the analysis can be used to recover an upper limit for the flux of the H$\alpha$ emission from the planet. \cite{Garcia99} discuss 
how the spectra of two unresolved companions can be recovered using the following equations
\begin{equation}
F_{*} = (1 / (1+r) + S(\lambda) / D)F_{total}(\lambda)
\end{equation}

\begin{equation}
F_{pl} = (r / (1+r) + S(\lambda) / D)F_{total}(\lambda)
\end{equation}
where {\it r} is the flux ratio of the two components in the continuum, S($\lambda$) is the centroid and {\it D} is their separation. As we detect no spectro-astrometric signature S($\lambda$) is replaced by the 3-$\sigma$ uncertainty measured for each value of $\lambda$ using Equation 1. To estimate the upper limit for the H$\alpha$ integrated flux we fit the H$\alpha$ line in the F$_{pl}$ spectrum. As we can only substitute the uncertainty for S($\lambda$), we cannot infer any information about the shape of the H$\alpha$ line profile in the spectrum of the companion. Therefore, it is assumed that shape of the star's and companion's H$\alpha$ line profile is the same. This is sufficient for estimating an upper limit for $\dot{M}_{acc}$ onto the planet. \cite{Kraus12} estimate the mass of LkCa 15b to be between 6~M$_{Jup}$ and 15~M$_{Jup}$. It was found that for a planet with this mass range the {\it r/(1+r)} term was negligible compared to the S($\lambda$)/D term. For example, considering the R and J band we find that our estimate lies between 10$^{-4}$ and 10$^{-2}$ while S($\lambda$) / D is on the order of 10$^{-1}$. Please refer to Appendix B for more details on how  {\it r} was estimated. Thus the upper limit on the flux from the planet was dependent on the uncertainty in the spectro-astrometric analysis and the distance between the star and the planet taken at 0$\farcs$1. The 3-$\sigma$ upper limit of the H$\alpha$ flux from the companion (averaged from all 6 spectra) is 1.0 $\times$ 10$^{-13}$~erg/s/cm$^{2}$ when a correction for A$_{v}$ = 1.2 \citep{Manara14} is applied. This value is used to estimate $\dot{M}_{acc}$ for the companion in Section 4.

\subsection{Analysis of NIR lines}
In Section 3.4 of their paper \cite{Kraus12} outline rejected alternative explanations for their observations. They discuss how SA of the Br$\gamma$ line could be used to investigate the possiblility that their 
observations are not of continuum emission from a planetary companion and circumplanetary dust but, are of line emission from gas in the disk gap. 
The Br$\gamma$, Pa$\gamma$, and Pa$\beta$ emission in each epoch is shown in Fig. \ref{lines} and no spectro-astrometric signature is detected for any of these lines. Also note how faint the Br$\gamma$ line is 
and the average S/N is found to be 6. This supports the arguments of \cite{Kraus12}. 
For the spectro-astrometric analysis the pipeline processed data was used as the NIR arm does not suffer from the spatial aliasing effects described in Section 3 and in the Appendix. Also
artefacts such as those seen in the H$\alpha$ line were not detected. Using the uncertainty in the spectro-astrometric analysis of the Pa$\gamma$ and Pa$\beta$ lines and the method outlined above the 3-$\sigma$ upper limit for the Pa$\gamma$ and 
Pa$\beta$ flux from the planetary companion is estimated at 7.0 $\times$ 10$^{-15}$ erg/s/cm$^{2}$ and 8.0 $\times$ 10$^{-15}$ erg/s/cm$^{2}$, respectively. These values have been corrected for extinction.
This flux is also used in Section 4 to place an upper limit on $\dot{M}_{acc}$ for the planet.


\section{Mass accretion in LkCa 15b}

Studies of mass accretion and $\dot{M}_{acc}$ onto planetary mass objects should help to distinguish their possible formation mechanisms. For example, \cite{Zhou14} carried out a study of $\dot{M}_{acc}$ onto three accreting objects, GSC 06214-00210b (GSC), GQ Lup b (GQ) and DH Tau b (DH). The mass range covered by these objects is 10~M$_{Jup}$ to 30~M$_{Jup}$ and they are among only a very small number of BDs / planets for which $\dot{M}_{acc}$ has been estimated. \cite{Zhou14} estimated $\dot{M}_{acc}$ using UV and optical photometry done with the 
Hubble Space Telescope (HST) and their method which follows \cite{Gullbring98} assumes that their objects accrete material from a disk in the same manner as pre-main sequence stars. They found their measured 
values of $\dot{M}_{acc}$ to be higher than expected when compared to the correlation between M$_{\star}$ and $\dot{M}_{acc}$, derived for YSOs and BDs, from UV excess measurements and other conventional methods for measuring $\dot{M}_{acc}$ \citep{Alcala14, Herczeg08}.  This correlation would apply to objects formed like stars (protostellar core fragmentation) and these results of \cite{Zhou14} could point to an alternative formation mechanism for GSC, GQ and DH. The authors briefly discuss formation by core accretion and disk instabilities. 
Furthermore, \cite{Zhou14} conclude that these high values of $\dot{M}_{acc}$ are consistent with the presence of massive disks around these objects 
from which they accrete, as their separations from their primaries (100-300 au) are too large to support accretion at such high rates directly from the primary.


We wish to add to the work of \cite{Zhou14} by deriving $\dot{M}_{acc}$ for LkCa15 b. To this end we use the upper limit on the flux from the H$\alpha$, Pa$\gamma$ and Pa$\beta$ lines to estimate the accretion luminosity (L$_{acc}$) in each line. The equation $\dot{M}_{acc}$ = 1.25 (L$_{acc}$ R$_{*}$) / (G M$_{*}$) is used \citep{Gullbring98, Hartmann98} and the results of \cite{Alcala14} are used to convert the luminosity of each line (L$_{line}$) to L$_{acc}$.  R$_{pl}$ is 
taken as 0.22~\Rsun\ and 0.34~\Rsun\ for masses of 6~M$_{Jup}$ 
and 15~M$_{Jup}$ respectively. R$_{pl}$ was estimated using the SETTL evolutionary models for 1Myr \citep{Allard03}. Our method is analogous to studies which use various accretion tracers to derive $\dot{M}_{acc}$ in low mass stars and BDs \citep{Alcala14, Whelan14a}, and therefore it is assumed here that the correlation between L$_{line}$ and $\dot{M}_{acc}$ for low mass pre-main sequence stars is applicable for planetary mass objects. Studies of $\dot{M}_{acc}$ using L$_{acc}$ estimated from L$_{line}$ have shown that the optimum way to estimate $\dot{M}_{acc}$ is to use several accretion tracers found in different wavelength regimes \citep{Rigliaco11, Rigliaco12}. In this way any spread in 
$\dot{M}_{acc}$ due to different tracers probing different regimes of accretion or having a jet/wind component can be overcome. This is our reason for using more than one accretion tracer here. However, as many of the accretion tracers found in the spectrum of LkCa 15 show strong variability (refer to Fig. \ref{lines}) we chose to limit our calculations to the lines with the least variability namely H$\alpha$, Pa$\gamma$ and Pa$\beta$. Here we are specifically referring to the presence of variable absorption features. For example in E1 the He I 1.083 $\mu$m line is double peaked but in E2 and E3 it is single peaked with a deep red-shifted absorption. The H$\gamma$ and H$\beta$ lines also show red-shifted absorption in E2.  Also H$\alpha$
is the line in which we are most likely to see a contribution from the planetary companion \cite{Close14}. It is found that for the 3-$\sigma$ upper limit on the flux, log($\dot{M}_{acc}$[H$\alpha$]) = -9.0 to -9.2, 
log($\dot{M}_{acc}$[Pa$\gamma$]) = -9.1 to -9.3 and log($\dot{M}_{acc}$[Pa$\beta$]) = -8.9 to -9.1. The range in $\dot{M}_{acc}$ for the individual lines comes from the estimated range in the mass of the companion. These estimates are in agreement with predictions for mass accretion rates from circumplanetary disks, which place $\dot{M}_{acc}$ for a 1-10~M$_{Jup}$ planet in the range 10$^{-9}$ to 10$^{-8}$~\Msun / yr \citep{Zhu15}. 

\begin{figure}[ht]
\centering
   \includegraphics[width=9.8cm, trim= 0.5cm 0cm 0cm 0cm, clip=true]{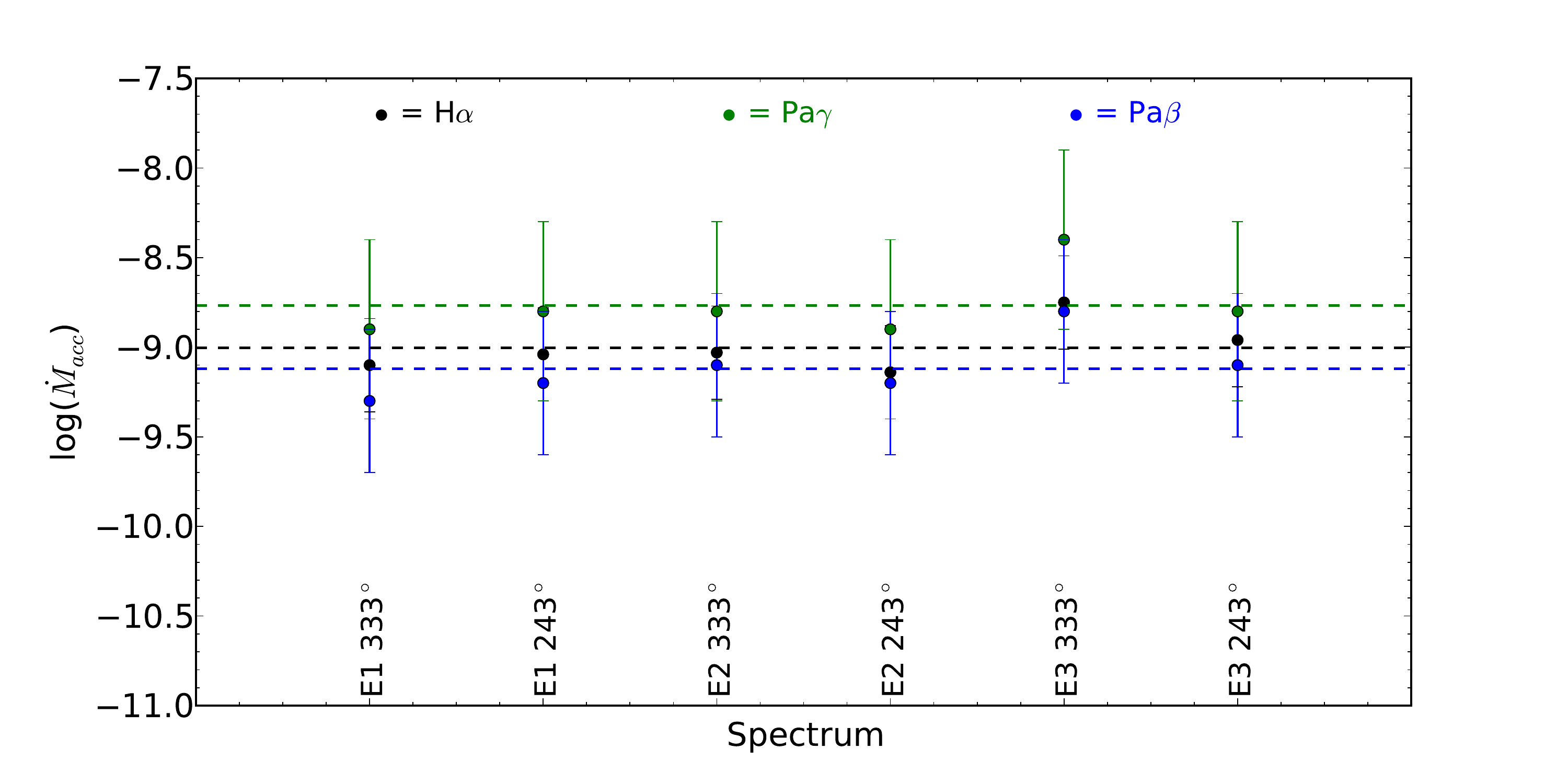}
     \caption{log($\dot{M}_{acc}$) calculated for all six spectra using the H$\alpha$, Pa$\gamma$ and Pa$\beta$ lines. The dashed lines are the mean values from each tracer.}
  \label{Macc}     
\end{figure}

To compare the estimated accretion rate of LkCa15b with that of the central object, we have also derived $\dot{M}_{acc}$ for LkCa15. In fact, \cite{Manara14} used the E1 spectra of LkCa 15 presented here, in their study of $\dot{M}_{acc}$ in YSOs with TDs.  
They make their estimates of $\dot{M}_{acc}$ by modeling the excess emission produced by disk accretion using a set of isothermal hydrogen slab emission spectra. 
As a check on their values of the accretion luminosity (L$_{acc}$) from the models, they also derive L$_{acc}$ using several different accretion tracers and the L$_{acc}$, 
line luminosity (L$_{line}$) 
relationships of \cite{Alcala14}. \cite{Manara14} report $\dot{M}_{acc}$ = 4 $\times$ 10$^{-9}$~\Msun / yr for LkCa 15. Here we estimate $\dot{M}_{acc}$ for LkCa 15 in the same manner as described above for LkCa 15b and again using the H$\alpha$, Pa$\gamma$ and Pa$\beta$ lines.  For LkCa 15
L$_{line}$ is corrected for extinction using A$_{v}$ = 1.2 and R$_{*}$ 
and M$_{*}$ are taken as 1.52~\Rsun\ and 1.24~\Msun\ respectively \citep{Manara14}. Our motivation is to check for any strong variability in $\dot{M}_{acc}$ for LkCa 15 between our three epochs of data. In Fig. \ref{Macc} log($\dot{M}_{acc}$) calculated using the H$\alpha$, Pa$\gamma$ and Pa$\beta$ lines from all six \xsh spectra are plotted. The dashed lines are the mean value of log($\dot{M}_{acc}$) in each line. Considering the three lines the mean value of $\dot{M}_{acc}$ is (1.3 $\pm$ 0.6) $\times$ 10$^{-9}$~\Msun / yr. This is the same order of magnitude but somewhat lower than the values reported in \cite{Ingleby13} and \cite{Manara14}.   Overall the variation in $\dot{M}_{acc}$ is not significant and is within the values reported by \cite{Costigan12} (0.37~dex) and \cite{Venuti14} (0.5~dex) in their studies of accretion variability in large samples of YSOs. 

\begin{figure*}[ht!]
\centering
   \includegraphics[width=15cm, trim= 0cm 0cm 0cm 0cm, clip=true]{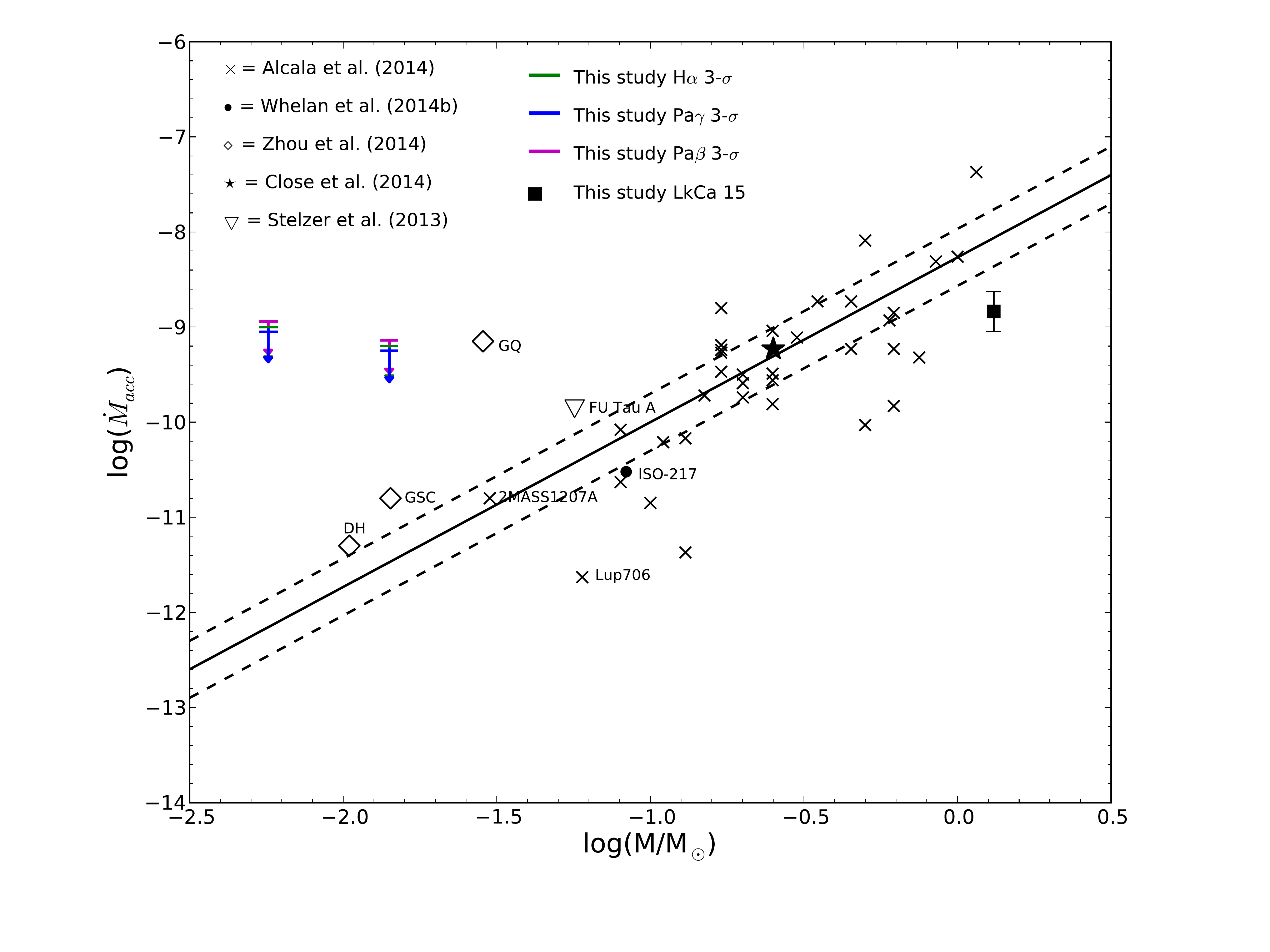}
     \caption{Comparing the upper limits for $\dot{M}_{acc}$ in LkCa 15b with values derived for accreting objects with masses within the range of low mass stars to planets. All the BDs and planets are marked with their names. For the objects from the \cite{Zhou14} study $\dot{M}_{acc}$ was estimated using UV and optical photometry. For the \cite{Close14} object $\dot{M}_{acc}$ was estimated from the H$\alpha$ emission detected with MagAO. For the rest of the objects $\dot{M}_{acc}$ was derived from \xsh data using the method outlined in Section 4. The majority of the sources are low mass stars and the values of $\dot{M}_{acc}$ are taken from \cite{Alcala14}. The black line represents the linear fit to the results of \cite{Alcala14} and it shows the correlation between M and $\dot{M}_{acc}$. The dashed lines represent the 1-$\sigma$ deviation from the fit. }
  \label{zhou}     
\end{figure*}

To put all these results into context we plot in Fig. \ref{zhou}, $\dot{M}_{acc}$ derived from \xsh data for a number of YSOs and BDs, as a function of mass. The results for the YSOs come from \cite{Alcala14} and the linear correlation marked by the black line comes from the fit to the \cite{Alcala14} results. 
Also shown are the results from \cite{Zhou14} and \cite{Close14}. \cite{Close14} use the Magellan Adaptive Optics system (MagAO) to detect a 0.25~\Lsun companion in the disk of the YSO HD142527. They estimate $\dot{M}_{acc}$ from the H$\alpha$ emission to be 5.9 $\times$ 10$^{-10}$ ~\Msun / yr and discuss how this techniques could be used to detect planets in their gas accretion phase. The names of the BDs and planets are marked on the figure. The estimates for $\dot{M}_{acc}$ in LkCa 15 and LkCa 15b are marked on Fig. \ref{zhou}. The upper limits we show here for LkCa 15b are comparable to $\dot{M}_{acc}$ for GQ and do not contradict the conclusions of \cite{Zhou14} that some planetary mass companions exhibit high mass accretion rates when compared to BDs and low mass YSOs. 

The central assumption of our study and that of \cite{Zhou14} is that magnetospheric accretion can also be used to describe accretion from a circumplanetary disk. However, it should be noted that it is not yet understood if magnetospheric accretion is the dominant mechanism by which planets accrete material from a disk. In their paper on observational signatures of accreting circumplanetary disks \cite{Zhu15} discuss a number of caveats to this assumption. If $\dot{M}_{acc}$ is high and the central accreting object has a weak magnetic field, accretion will occur via a boundary layer rather than along magnetic field lines. In this scenario strong optical and UV emission from the boundary layer could be expected. Furthermore, if magnetic fields are high enough to truncate the inner disk and for magnetospheric accretion to occur, it is not clear if heating mechanisms are sufficient to produce strong line emission such as what is seen in pre-main sequence stars. As more and more planetary mass objects are detected a clearer understanding $\dot{M}_{acc}$ at planetary masses should emerge. Future studies of circumplanetary disks with the James Webb Space Telescope (JWST) for example, should also answer many questions.

\section{Summary and conclusions}
Here multi-wavelength spectroscopic observations of the transitional object \Lk obtained with \xsh are presented. The goal was to investigate the possibility of using spectro-astrometry to detect planetary mass objects in transitional disks. It is planned that the data presented here will be part of an extensive examination of the variability of \Lk and further data will be obtained in winter 2014-2015. Our results and conculsions can be summarised as follows. 

X-Shooter is not the best choice for a spectro-astrometric study due to the artefacts which can be introduced by the processing of the data and by instrumental distortions of the PSF. The artefacts were primarily a problem for the VIS and UVB arms and for the broadest brightest lines, i.e H$\alpha$. X-Shooter is a complex instrument with
 flexures both from thermal and gravity effects (due to its Cassegrain position), in
 addition to the curved and tilted spectral format (which necessitates elaborate
 data processing). Instruments which do not require such a high level of data processing to produce the final spectrum would be a better choice for spectro-astrometry. 

Although no spectro-astrometric signatures were detected for any of the accretion lines the uncertainty in the analysis could be used to put an upper limit on the flux from the planetary companion, in each line investigated. This was done for the H$\alpha$, Pa$\gamma$ and Pa$\beta$ lines. The derived fluxes when corrected for extinction were used to place an upper limit on the planetary companions mass accretion rate. It was found that log($\dot{M}_{acc}$) = -8.9 to -9.3 for the mass of the companion between 6 M$_{Jup}$ and 15 M$_{Jup}$, respectively. This is in agreement with estimates of mass accretion rates from circumplanetary disks. However, a more accurate estimate is needed before any conclusions can be drawn about the formation of LkCa 15b.

Overall we conclude that spectro-astrometry may not be the most efficient mechanism for detecting young planets in transitional disks. Even without the difficulties caused by spectro-astrometric artefacts, observations would have to be made when {$\it F_{*}$} / {$\it F_{pl}$}  is favorable for a detection and sub-milliarcsecond accuracy would be needed. However, the results we present here do show that it should be possible to detect LkCa 15b in H$\alpha$ with high angular resolution imaging techniques for example, with VLT / SPHERE or MagAO.  Taking the maximum value of the H$\alpha$ flux from the star as 1.54 $\times$ 10$^{-12}$~erg/s/cm$^{2}$ (measured for the E3 333$^{\circ}$ spectrum), the upper limit on the H$\alpha$ flux from the companion would give a contrast of $\sim$ 3.4~mag between the star and planet. It is feasible to detect this with SPHERE, which provides a contrast of $\sim$ 6.5~mag at 80~mas
in 1 hour of exposure time, in the H$\alpha$ narrow band filter. This is still feasible even if we take the flux from the planet as being 50 times less than the upper limit reported here. These observations could then be used to better estimate $\dot{M}_{acc}$. Furthermore, MagAO has already been able to spatially resolve a
companion in H$\alpha$ within the gap of the transitional disk of the Herbig Ae/Be star HD142527 \citep{Close14}. Here $\Delta$mag = 6.33 $\pm$ 0.20~mag at the H$\alpha$ line. A detection of LkCa 15b in H$\alpha$ would allow the estimate for $\dot{M}_{acc}$ given here to be greatly refined.

\acknowledgements{We thank the referee M. Takami for his very helpful comments. We thank P. Goldoni for discussions on the aliasing problems
in long-slit spectroscopy in general and for X-Shooter in particular.
We thank S. Moehler, V. D'Oricio, P. Goldoni and A. Modigliani for their help
with the X-shooter pipeline, and F. Getman and G. Capasso for the
installation of the different pipeline versions at Capodimonte. We acknowledge support from the Faculty of the European Space Astronomy Centre (ESAC). E.T. Whelan acknowledges financial support from the Deutsche Forschungsgemeinschaft through the Research Grant Wh 172/1-1. This research has also been funded by Spanish grant AYA2012-38897-C02-01. J.M. Alcal{\'a} acknowledges financial support from the PRIN INAF 2013
 “Disks, jets and the dawn of planets” }

{}


\begin{appendix}

\section{Spectro-astrometric analysis of the pipeline processed data.}

\begin{figure*}[ht!]
\centering
   \includegraphics[width=7cm, angle=0, trim= 0cm 0cm 0cm 0cm, clip=true]{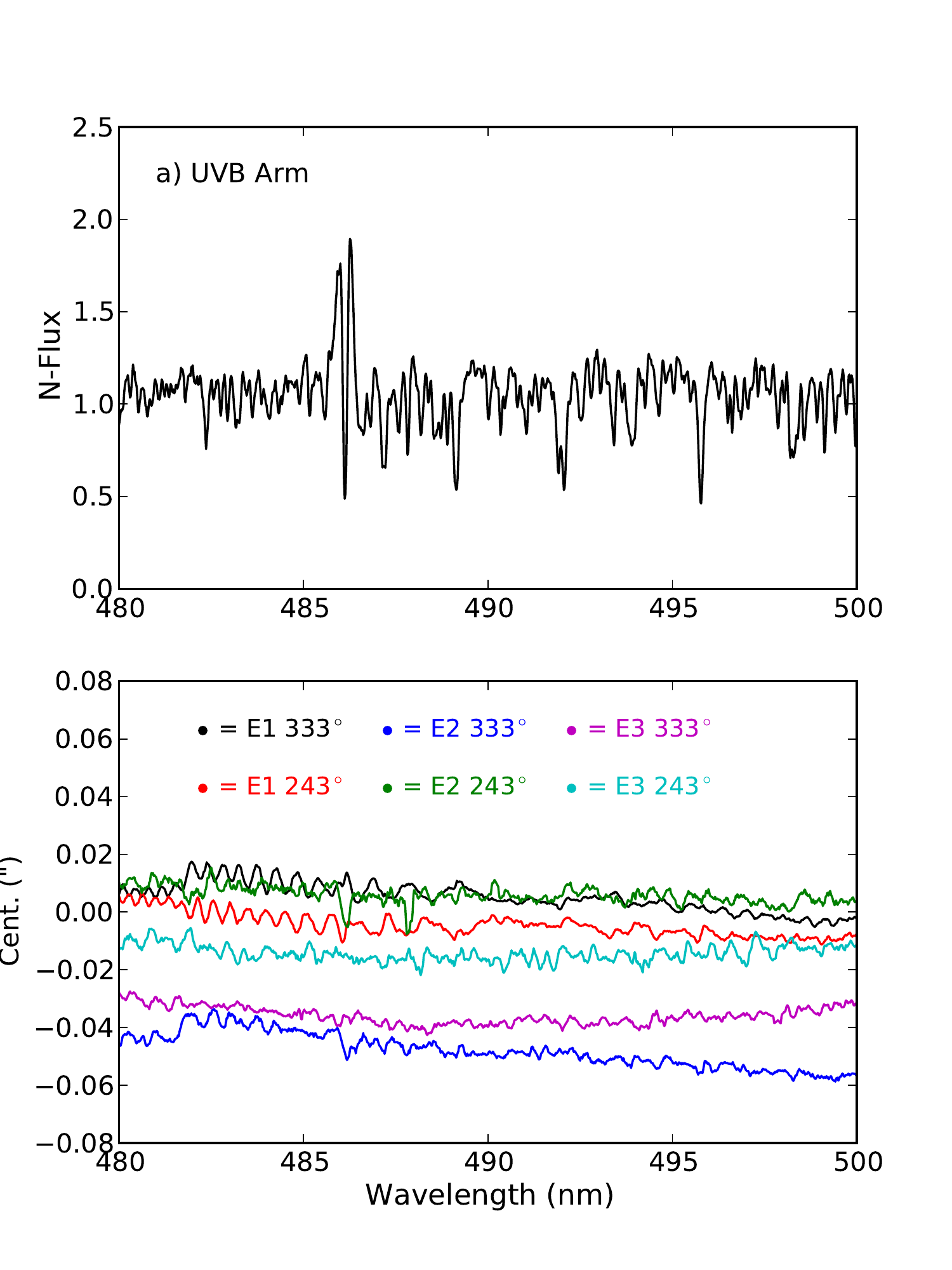}
   \includegraphics[width=7cm, angle=0, trim= 0cm 0cm 0cm 0cm, clip=true]{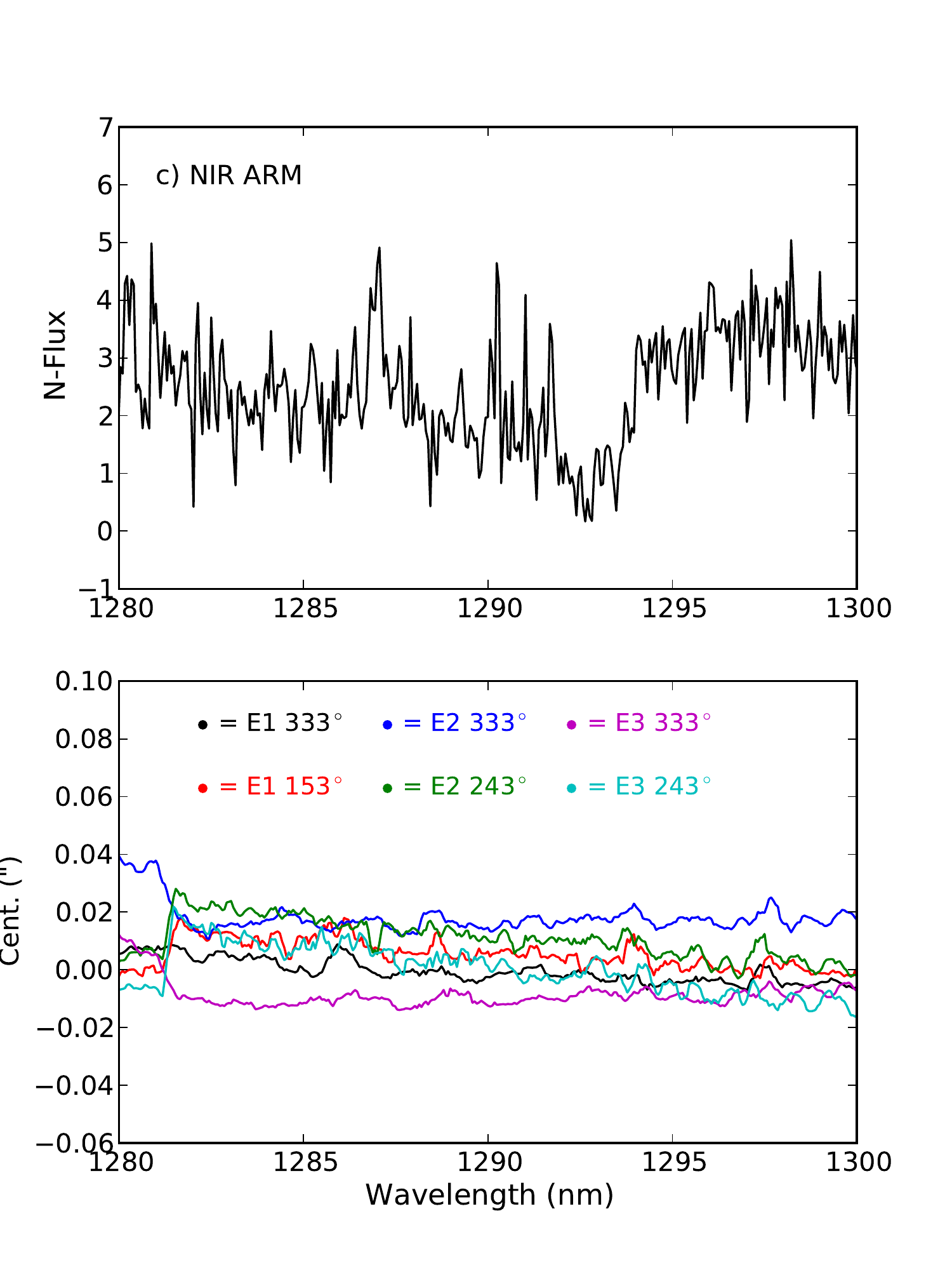}
   \includegraphics[width=7cm, angle=0, trim= 0cm 0cm 0cm 0cm, clip=true]{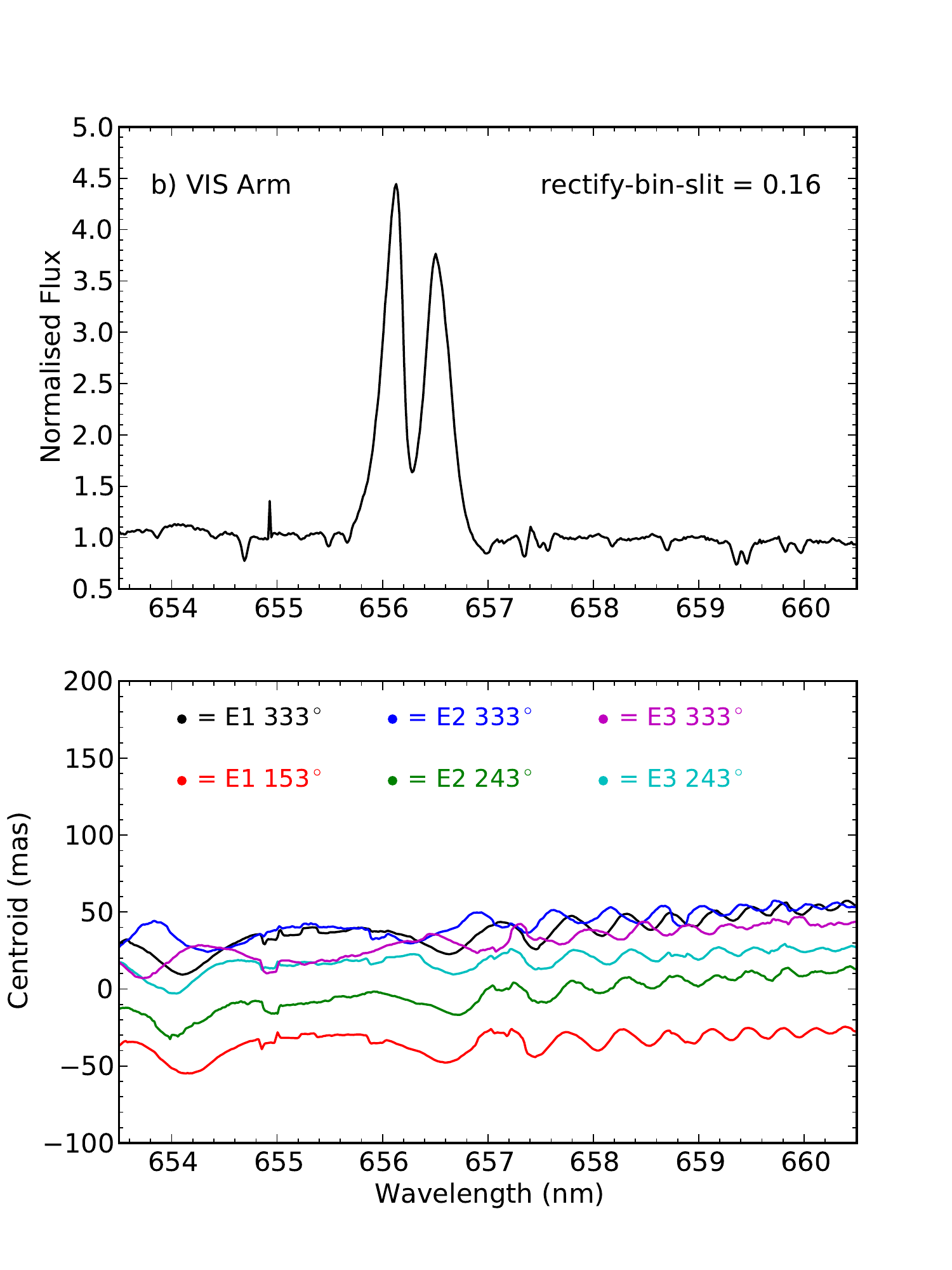}
      \includegraphics[width=7cm, angle=0, trim= 0cm 0cm 0cm 0cm, clip=true]{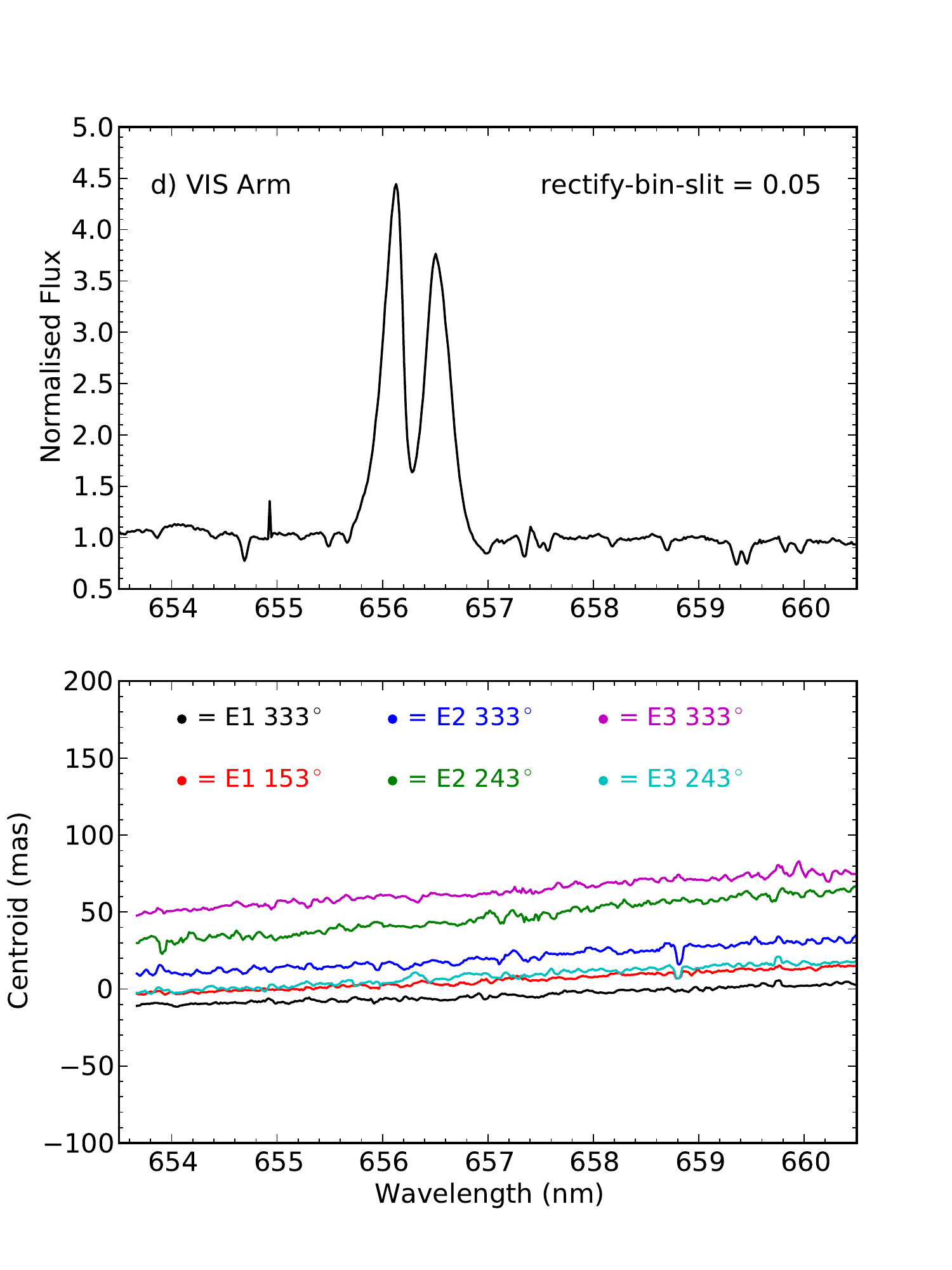}
     \caption{Spectro-astrometric analysis of a section of the UVB, VIS and NIR spectra. Note the small-scale 
     oscillations in the UVB and VIS spectra. This is caused by spatial aliasing and the effect is minimised by reducing the bin
     used to interpolate the pixels along the slit.} 
  \label{SA_wiggles}     
\end{figure*}

The full UVB, VIS and NIR in each epoch and at each PA were analysed using spectro-astrometry. In the VIS arm and the brightest parts of the UVB, an oscillating pattern with an amplitude of $\sim$ 10~mas was seen. To illustrate this better the position spectra for all six spectra (three epochs x 2 PAs) for the regions of the H$\beta$ (panel a), H$\alpha$  (panel b) and Pa$\beta$ lines (panel c) are presented in Fig. \ref{SA_wiggles}. 
It is concluded that this signal is caused by {\it spatial aliasing} introduced by the data reduction. This signal is not present in the raw data (refer to Fig. \ref{SA_Ha}). 
Four to five pixels are normally necessary to adequately
sample the FWHM of a spatial profile given by the seeing. If under sampled the effects of spatial ``aliasing" appear, which show-up as spatial oscillations
in the 1D spectra and in their corresponding 2D frames. The problem arises here due to the rebinning of the spectra from the physical pixel space (x,y) to the virtual pixels (wavelength, slit-scale) and is enhanced for high S/N and in the wavelength range of highest instrumental sensitivity, which in this case is the VIS arm. 
The sensitivity to S/N is the reason why we do not see this effect in the NIR arm 
and only in the brightest parts of the UVB arm. Also this effect will be enhanced in good seeing conditions. For example, note that in the UVB arm the effects are worse in the E1 spectra which 
has considerably better seeing than the E2 or E3 data (see Table 1). 

From Fig. \ref{SA_wiggles}, it is clear that this effect will have
important consequences for any spectro-astrometric analysis. Being intrinsic to the interpolation method along the slit, 
the problem cannot be
totally removed, however its effect can be reduced. This was done by using a smaller bin
than the standard one to interpolate the pixels along the slit. Therefore, in order
to minimize the aliasing effect on the spectro-astrometric analysis, we have used
a value of 0\farcs05 as input for the binning parameter ``rectify-bin-slit'' instead of
the standard value of 0\farcs16. As is seen in Fig. \ref{SA_wiggles}, panel d, this approach solved the spatial aliasing problem in the 
VIS arm. However, it should be considered that the re-binning in order to solve the aliasing problem could alter any real spectro-astrometric signal.

\section{Continuum flux ratio of the LkCa 15 it's companion}

To estimate the continuum flux ratio of LkCa 15 and it's planetary companion ({\it r}) we need to derive the flux of the star and planet at the given wavelength.  In the case of the star and the H$_{\alpha}$ line, we have considered the faintest Johnson R-magnitude from Grankin et al. (2007), since it represents the quiescent state of LkCa15 (R=11.97 mag, or M$_{R}$= 6.23\,mag for a distance of 140\,pc).  In the case of the planet, we have estimated M$_{R}$ for a 15 and a 6\,M$_{Jup}$ object using the SETTL evolutionary tracks (Johnson system) for 1\,Myr. We obtain values of 13.486 and 16.47\,mag, respectively, and a difference in magnitudes of 7.256\,mag and 10.24\,mag with respect to the star.  We estimate $r$ values of $\sim$10$^{-3}$ and 8$\times$10$^{-5}$ for 15 and 6\,M$_{Jup}$. Therefore, for a planet in this mass range, the$r/(1+r)$ term is negligible. For the Paschen lines, we have repeated the same procedure using the 2MASS J-mag of the star (J=9.42$\pm$0.02\,mag), and the 1\,Myr SETTL models in the 2MASS system to estimate the magnitude of the planet. We estimate a difference in magnitudes, $\Delta J$, between the planet and the star of 5.09\,mag and 6.845\,mag for a 15 and a 6 M$_{Jup}$ object, respectively. The $r$ values are $\sim$ 9$\times$10$^{-3}$ and $\sim $2.10$^{-3}$, so that the $r/(1+r)$ term is also negligible.

The above ratio was estimated assuming only photospheric emission, that is, not considering accretion.
 Firstly we stress that the contrast has been calculated in the
 R-band where the continuum excess due to accretion onto a planetary
 mass object is much less than in the blue (see Fig.2 by \cite{Zhou14}). From the slab models shown in Fig 2 of \cite{Zhou14} one can see that the continuum excess in the R-band is in most cases just a fraction of
 the photospheric emission. {\bf The exception is GQ Lup where the excess continuum emission is several times the photospheric emission at a wavelength close to the H$\alpha$ line. Taking the example of GQ Lup as the extreme case we place a factor of 10 as the upper limit on the ratio between the continuum excess emission and the photospheric emission. Increasing $r$ by a factor of 10 increases the range we estimate for $r$ to between $\sim$10$^{-2}$ and 8$\times$10$^{-4}$. Thus $r$ is still negligible compared to all values of S($\lambda$) / D, and we find that this increase does not change our estimate for the upper limit for the H$\alpha$ line flux from the planet. }The continuum excess emission due to accretion onto the planet, is also negligible in the NIR (see again Fig.2 in \cite{Zhou14}). Therefore, the contrast is not affected.

\end{appendix}

\end{document}